\documentclass[twocolumn,showpacs,amsmath,amssymb]{revtex4}
\input{epsf}

\usepackage{graphicx}
\usepackage{array}
\usepackage{dcolumn,longtable}

\usepackage{rotating,booktabs}
\usepackage{booktabs,threeparttable}

\begin{document}

\title{\bf
Convergence of the multipole expansion of the polarization interaction
}
\author{Yong-Hui Zhang$^{1,2}$, Li-Yan Tang$^{1}$, Xian-Zhou Zhang$^{2}$, and J. Mitroy$^{3}$}

\affiliation {$^1$State Key Laboratory of Magnetic Resonance and
Atomic and Molecular Physics, Wuhan Institute of Physics and
Mathematics, Chinese Academy of Sciences, Wuhan 430071, P. R. China}

\affiliation {$^{2}$Department of Physics, Henan Normal University,
XinXiang 453007, P. R. China}

\affiliation {$^{3}$School of Engineering, Charles Darwin
University, Darwin NT 0909, Australia}

\date{\today}

\begin{abstract}
The multipole expansion of the polarization interaction between a charged
particle and an electrically charged particle has long been known to be
asymptotic in nature, i.e. the multiple expansion diverges at any finite
distance from the atom. However, it is shown that the multipole expansion of
the polarization potential of a confined hydrogen atom is absolutely convergent
at a distance outside the atoms confinement radius.
It is likely that the multipole expansion of the dispersion interaction of
two confined atoms will also be absolutely convergent provided the internuclear
separation of the two atoms is sufficiency large to exclude any overlap between
the electron charge clouds of the two atoms.

\end{abstract}

\pacs{34.20.Cf, 31.15.ap } \maketitle

\section{Nature of the polarization expansion}

The long-range interactions between a charged particle and an atom
is usually described by a polarization potential.  This interaction
is constructed by making a multipole expansion of the two-particle
coulomb interaction leading to the following long-range expression
for the second-order adiabatic polarization potential,
\begin{equation}
V_{\rm pol}(R) \sim - \sum_{\ell=1}^{\infty} V^{(\ell)}_{\rm pol}(R)
= - \sum_{\ell=1}^{\infty} \frac{\alpha_{\ell}}{2R^{2+2\ell}} \ .
\label{Vpol1}
\end{equation}
In this expression, the $\alpha_{\ell}$ are the static multipole
polarizabilities. The most important term arising from $\ell = 1$,
the static dipole polarizability is sometimes called the
polarizability.  Equation (\ref{Vpol1}) has been long known to be an
asymptotic expansion.  The polarization potential given by
Eq.~(\ref{Vpol1}) eventually diverges as $\ell$ increases at any
finite value of $R$ \cite{roe52a,dalgarno56a}.

Formal issues of a similar nature also lead to a problem in the multipole
expansion of the atom-atom dispersion interaction.  Dalgarno and Lewis \cite{dalgarno56a}
originally showed the multipole expansion of the H-H second-order dispersion
interaction,
\begin{equation}
V_{\rm disp}(R) \sim - \sum_{n=1}^{\infty} \frac{C_{4+2n}}{R^{4+2n}}
\ , \label{Vdisp1}
\end{equation}
also diverges in the Unsold mean energy approximation \cite{unsold27a}, as
$n \to \infty$ at any finite $R$.  Sometime later it was shown that the exact
multipole series for the H-H interaction was also asymptotic
\cite{young75a,ahlrichs76a}.

Recently, B-spline basis sets have become very popular for solving
atomic structure and collisions problems
\cite{johnson88a,sapirstein96a,shi00a,kang06a,bachau01a}. Such
methods  often impose the boundary condition that the wave function
is zero at some finite radial distance, i.e. effectively solving the
Schrodinger equation in a sphere of finite volume.  One aspect of
bounding the wave function is that the reaction of the wave function
to an electric field also becomes bounded in space.  This results in
a multipole expansion of the polarization interaction that can be
shown to be absolutely convergent provided that the distance from
the atom is larger than the confining radius of the atom.  The
imposition of a $V = \infty$ boundary will result in a multipole
expansion that is mathematically well behaved at the cost tolerating
a small error in the size magnitude of the polarization potential.

In this manuscript the Unsold approximation is applied to estimate
the multipole polarizabilities of a confined H atom.  The resulting
polarization expansion is then shown to be absolutely convergent
when the radius is larger than the boundary of the confinement
potential.  This is supplemented by finite radius B-spline
calculations of the hydrogen polarizabilities for a box size of $R_0
= 14$ $a_0$.  The results of these explicit calculations confirm the
Unsold analysis and it is shown that the error in the polarization
potential arising from the use of finite radius polarizabilities
does not exceed 3 part in $10^{6}$.

\section{Polarization interaction for a confined hydrogen atom}

The static multipole polarizabilities of any spherically symmetric
state are defined
\begin{equation}
\alpha_{\ell} = \sum_i \frac{f^{(\ell)}_{0i}}{(\Delta E_{0i})^2 }
\label{alphal}
\end{equation}
where $\Delta E_{0i}$ is the excitation energy from state $0$ to
state $i$, the sum implicitly includes the continuum, and
$f^{(\ell)}_{0i}$ is the oscillator strength of multipole $\ell$
connecting the state 0 to the excited state $i$.  The
$f^{(\ell)}_{0i}$ are defined \cite{yan96a,mitroy03f}
\begin{equation}
f^{(\ell)}_{0i} =  \frac {2 |\langle \psi_0 \parallel r^{\ell} {\bf
C}^{\ell}({\bf \hat{r}} ) \parallel \psi_{i}\rangle|^2 \Delta
E_{0i}} {(2\ell+1) }  \ . \label{fvaldef}
\end{equation}
In this expression ${\bf C}^{\ell}$ is the spherical tensor of rank
$\ell$.

Consider the oscillator strength sum rule
\begin{equation}
S^{(\ell)}(0) = \sum_i f^{(\ell)}_{0i} \label{S0}
\end{equation}
This satisfies the identity
\begin{equation}
S^{(\ell)}(0) =  N \ell \langle r^{2\ell-2} \rangle
\label{fsuml}
\end{equation}
\cite{rosenthal74a} where $N$ is the number of electrons and
$\langle r^{2\ell-2} \rangle$ is a radial expectation value of
the ground state wave function.  This expression
reduces to the well known Thomas-Reiche-Kuhn sum rule $S^{(1)}(0)  = N$
for $\ell = 1$.  For the hydrogen atom ground state, the sum rule is
\begin{equation}
S^{(\ell)}(0) =  \frac{\ell(2\ell)!}{2^{2\ell-1}}
\label{fsumH}
\end{equation}
The exact value of the multipole polarizabilities for the hydrogen
atom ground state can be written \cite{dalgarno55a}
\begin{equation}
\alpha_{\ell} = \frac{(2\ell+2)!(\ell+2)}{\ell(\ell+1)2^{2\ell+1}}
\end{equation}

For a confined atom, the radial expectation value in Eq.~(\ref{fsuml}) will
be bounded by the radius of the $B$-spline box, $R_0$, i.e.
\begin{equation}
\langle  r^{2\ell-2} \rangle \le  R_0^{2\ell-2}.
\label{radial}
\end{equation}
Note that the correctness of Eq.~(\ref{fsuml}) for calculations in a
finite box has been validated to 20 significant figures.

Eq.~(\ref{radial}) shows the factorial growth in the sum-rule, Eq.~(\ref{fsumH})
is eliminated once the wave function is confined.
For the hydrogen atom ground state one has
\begin{equation}
S^{(\ell)}(0) \le \ell R_0^{2\ell-2}
\label{fsumH2}
\end{equation}

Confining a hydrogen atom also impacts the energies of the states.
So the energies used to calculate the oscillator strengths and the
energies used in the energy denominators of Eq.~(\ref{alphal}).  As
$\ell$ increases in size, the centrifugal potential will tend to
dominate the coulomb potential for $r \le R_0$.  For $\ell >
\sqrt{2ZR_0}$ the total potential is repulsive and increases
monotonically for decreasing $R$.  When this occurs, the minimum
excitation energy will be larger than the potential at $R = R_0$,
i.e $\Delta E = \ell(\ell+1)/2R_0^2 - 1/R_0 + 0.5$. For sufficiently
large $\ell$ one can write $\Delta E \approx \ell({\ell+1})/2R_0^2$.

Making the Unsold mean energy approximation \cite{unsold27a}, and
setting the mean excitation energy to the minimum value, leads to
the following large $\ell$ approximate expression for the hydrogen
polarizabilities
\begin{equation}
\alpha_{\ell} \approx \frac{ 2 R_0^{2\ell} } { (\ell+1) }
\end{equation}
The ratio of two successive terms in Eq.~(\ref{Vpol1}),
\begin{equation}
T =  \frac{ V^{(\ell+1)}_{pol}(R)  }{ V^{(\ell)}_{pol}(R) }
 =  \frac{ (\ell+1) R_0^2  }{ (\ell+2) R^2 }
\end{equation}
This ratio is less than one for $R_0/R$ less than 1.  Therefore the
polarization series is absolutely convergent as long as $R$ is greater
than $R_0$.

This absolute convergence has been substantiated by calculations of
the multipole polarizabilities of the hydrogen ground state.
Calculations were performed using a B-spline basis with the outer
boundary set to $R_0 = 14$ $a_0$ and results are reported in Table
\ref{tab1}. The static polarizabilities of hydrogen have been
previous calculated using B-splines basis sets \cite{bhatti03a} but
the intent of the present work is different from the earlier
investigations.

The radial wave function is written as B-splines
\begin{equation}
\Psi(r) = \sum_i c_i B_i(r) \label{BB}
\end{equation}
where $\Psi(r)$ is normalized such that
\begin{equation}
\int_{0}^{\infty} |\Psi(r)|^2 \ r^2 \ dr = 1 . \ \ \\
\end{equation}
The boundary conditions are such that the first B-spline is finite at the origin.

\begin{figure}[tbh]
\centering{
\includegraphics[width=8.4cm,angle=0]{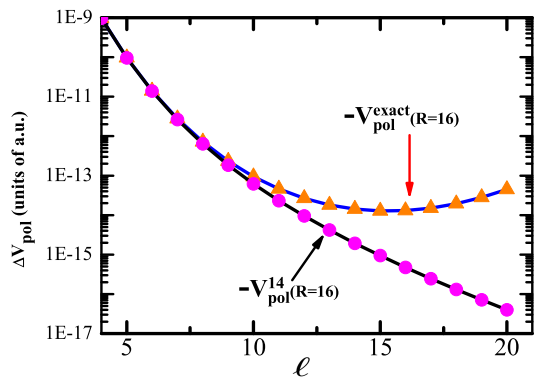}
} \caption[]{ \label{fig1} The magnitude of the contribution to
$V_{\rm pol}(R=16)$ from each successive multipole of
Eq.~(\ref{Vpol1}).  Results for $R_0 = 14$ $a_0$ and the unconfined
atom are depicted. }
\end{figure}

\begin{table*}
\caption{\label{tab1} The multipole polarizabilities (in a.u.) of
the hydrogen atom ground state.  Both exact and B-spline
polarizabilities calculated in a sphere of radius $R_0 = 14$ $a_0$
are given.  The polarization potential at $R = 16$ $a_0$ is computed
with Eq.~(\ref{Vpol1}) including terms of successively larger $\ell$
for both sets of polarizabilities.  The column $E_{\rm lowest}$
gives the lowest B-spline energies for each value of $\ell$. }
\begin{ruledtabular}
\begin{tabular}{ccllll}
\multicolumn{1}{c}{$\ell$}  & \multicolumn{1}{c}{$E_{\rm lowest}$} &
\multicolumn{1}{c}{$\alpha_{\ell}^{\infty}$} &
\multicolumn{1}{c}{$\alpha^{14}_{\ell}$: B-spline} &
\multicolumn{1}{c}{$10^{-5} \times V^{\infty}_{\rm pol}(R=16)$}
& \multicolumn{1}{c}{$10^{-5} \times V^{14}_{\rm pol}(R=16)$}   \\
\hline
1  &$-$0.124540597990     & 4.500000000000000                  &4.499992173038246                       &$-$3.433227539062500   &$-$3.433221567564579\\
2  &$-$0.043113470410     & 15.00000000000000                  &14.99962361522554                       &$-$3.477931022644042   &$-$3.477923929432083\\
3  &0.011642665533        & 131.2500000000000                  &131.2214665586345                       &$-$3.479458973743021   &$-$3.479451548358086\\
4  &0.065428512840        & 2.126250000000000$\times$$10^{3}$  &2.123507410263223$\times$$10^{3}$       &$-$3.479555664398503   &$-$3.479548114295026\\
5  &0.122659506743        & 5.457375000000000$\times$$10^{4}$  &5.426703060742913$\times$$10^{4}$       &$-$3.479565358643910   &$-$3.479557754056124\\
6  &0.184625743002        & 2.027025000000000$\times$$10^{6}$  &1.988753009195520$\times$$10^{6}$       &$-$3.479566765175051   &$-$3.479559134030737\\
7  &0.251827759083        & 1.026181406250000$\times$$10^{8}$  &9.740932380195944$\times$$10^{7}$       &$-$3.479567043322079   &$-$3.479559398059238\\
8  &0.324490527950        & 6.784199296875000$\times$$10^{9}$  &6.020582918170891$\times$$10^{9}$       &$-$3.479567115152583   &$-$3.479559461804639\\
9  &0.402722627126        & 5.671590612187500$\times$$10^{11}$ &4.473394110827833$\times$$10^{11}$      &$-$3.479567138609732   &$-$3.479559480306164\\
10 &0.486577112031        & 5.846894322018750$\times$$10^{13}$ &3.838766302059065$\times$$10^{13}$      &$-$3.479567148055899   &$-$3.479559486508025\\
11 &0.576078316293        & 7.284255842848359$\times$$10^{15}$ &3.685245972773515$\times$$10^{15}$      &$-$3.479567152652911   &$-$3.479559488833742\\
12 &0.671234886264        & 1.078630192114083$\times$$10^{18}$ &3.862153915918611$\times$$10^{17}$      &$-$3.479567155311939   &$-$3.479559489785837\\
13 &0.772046602041        & 1.872193833455160$\times$$10^{20}$ &4.337864044592644$\times$$10^{19}$      &$-$3.479567157114797   &$-$3.479559490203558\\
14 &0.878508149199        & 3.764357734467175$\times$$10^{22}$ &5.150165059573718$\times$$10^{21}$      &$-$3.479567158530791   &$-$3.479559490397286\\
15 &0.990611293536        & 8.679197301530880$\times$$10^{24}$ &6.397025772360861$\times$$10^{23}$      &$-$3.479567159806084   &$-$3.479559490491281\\
16 &1.108346175117        & 2.274460234018827$\times$$10^{27}$ &8.248030398257013$\times$$10^{25}$      &$-$3.479567161111560   &$-$3.479559490538623\\
17 &1.231702096001        & 6.722293580544535$\times$$10^{29}$ &1.097317979742808$\times$$10^{28}$      &$-$3.479567162618750   &$-$3.479559490563226\\
18 &1.360668006907        & 2.225432980085533$\times$$10^{32}$ &1.499328767223332$\times$$10^{30}$      &$-$3.479567164567810   &$-$3.479559490576357\\
19 &1.495232809884        & 8.201833248105232$\times$$10^{34}$ &2.096237726358987$\times$$10^{32}$      &$-$3.479567167373771   &$-$3.479559490583528\\
20 &1.635385546034        & 3.346738528714939$\times$$10^{37}$ &2.990044876206528$\times$$10^{34}$      &$-$3.479567171846294   &$-$3.479559490587524\\
\end{tabular}
\end{ruledtabular}
\end{table*}

The polarizabilities listed in Table \ref{tab1} show the increasing difference
between the the B-spline calculation of $\alpha_{\ell}$ from the exact
calculation at the higher values of $\ell$.  The polarizabilities
are written as $\alpha^{R_0}_{\ell}$ where $R_0$ specifies the outer
radius of the B-spline box.   The B-spline $\alpha^{R_0}_{\ell}$
increases less rapidly than the exact value,
$\alpha^{\infty}_{\ell}$, and at $\ell = 20$ is 1000 times smaller
than the exact polarizability.

The convergence of the $B$-spine polarization potential at $R = 16$
$a_0$ is also shown in Figure \ref{fig1}.  The contributions of each
individual multipole in Eq.~(\ref{Vpol1}), i.e. $V^{14}_{\rm
pol}(R)$, are plotted as a function of $\ell$.  The divergence in
the polarization series is apparent as the increments start to
increase for $\ell \ge 15$. However, the increments steadily
decrease as $\ell$ increases for the multipole expansion for the
B-spline $V_{\rm pol}^{14}(R=16)$ providing a computational
demonstration of the absolute convergence of the B-spline expansion.

The imposition of the boundary at $R_0 = 14$ $a_0$
has resulted in a polarization potential that is 0.0003$\%$
smaller in magnitude at $R = 16$ $a_0$ than the expansion using
the exact polarizabilities.  Most of this difference arises
from the dipole and quadrupole terms in the multipole
expansion.

The demonstrated convergence of the B-spline expansion has some
relevance to the earlier analysis by Brooks \cite{brooks52a}.   He stated
that the origin of the multipole series divergence was the use
of the expansion
\begin{equation}
V = \frac{1}{|{\mathbf r}_1 - {\mathbf r}_2|} = \sum_{k=0}^{\infty}
\frac{ r_{<}^{k} } {r_{>}^{k+1} }
  {\mathbf C}^k({\hat {\mathbf r}}_1)\cdot{\mathbf C}^k({\hat {\mathbf r}}_2)
\label{r12}
\end{equation}
beyond its region of validity.  This analysis recommended that all
radial matrix elements involved in the evaluation of the
polarizability have their radial integrations limited to a finite
value, e.g. $X$, which would a lower bound on the region for which
one writes the polarization potential using Eq.~(\ref{Vpol1}).  The
interaction, Eq.~(\ref{r12}) is effectively reduced to
\begin{equation}
V = \sum_{k=0}^{\infty}
\frac{ r^k } {X^{k+1} }
  {\mathbf C}^k({\hat {\mathbf r}})\cdot{\mathbf C}^k({\hat {\mathbf X}}) \ \ \ \ (r \le X)
\end{equation}
where $X$ is a lower bound to the radial coordinate used to evaluate
$V_{\rm pol}$. The resulting multipole expansion was then shown to
be convergent.  Brookes concluded that the divergence in the
polarization expansion was caused by using Eq.~(\ref{r12}) in
regions of space where it is not correct.  This statement is
correct, but cannot be used to imply that the evaluation of the
second-order polarization interaction using $|{\mathbf r}_1 -
{\mathbf r}_2|^{-1}$ without making the multipole expansion will
lead to a convergent expansion in powers of $R^{-1}$.  Indeed, it
has been shown that by an exact analysis in the Unsold approximation
that the power series expansion $R^{-1}$ of the hydrogen atom
polarization potential is asymptotic \cite{roe52a,dalgarno56a}.
However, using a wave function that is limited in radial extent, or
using an interaction that is limited in radial extent
\cite{brooks52a} will lead to a convergent expansion of the
polarization interaction.

\section{Conclusion}

It has been demonstrated that the multipole expansion of a confined
hydrogen atom polarization potential leads to an inverse power
series that is absolutely convergent provided the distance from the
nucleus is larger than the box radius. While the result is rigorous
for hydrogen, one can reasonably assert that the result can be
applied to any atom. This result has implications that go beyond the
second-order perturbation expansion of the polarization interaction.
For example, the Stark effect is known to be divergent with respect
to order of perturbation theory \cite{kato76a}.  It is interesting
to speculate whether bounding the wave function will also remove
this divergence.

The multipole expansion of the atom-atom dispersion interaction is also known
to be asymptotic in nature \cite{dalgarno56a,young75a,ahlrichs76a}.  It is
reasonable to speculate that this expansion would also be absolutely convergent
for two confined atoms provided the internuclear distance was larger than the
sum of the two box sizes.  It is interesting to note that Jansen \cite{jansen00a}
had suggested that the $V_{\rm disp}$ multipole expansion would converge provided the
basis set used
in the evaluation of the oscillator strength sum rule was restricted in dimension
and range \cite{jansen00a}.  The numerical results associated with the work
of Jansen were rather restricted in scope.  Koide had devised a momentum
space treatment that resulted in each dispersion parameter being multiplied
by a radial cutoff of factor \cite{koide76a}.  Although it could be shown
that the resulting dispersion interaction was convergent at any finite $R$
as $n \to \infty$ the method has only ever been applied to a few terms
of the H-H dispersion interaction.

One aspect of the present analysis should be emphasised.  Confining the
atomic wave function to a finite region of space does constitute
the imposition of an artificial constraint upon the wave function.
However, such constraints are routinely applied in calculations
of atomic structure based on B-spline basis sets.  These descriptions
of atomic structure will automatically have no formal issues with respect
to the convergence of the multipole expansion of the polarization
interaction.

\begin{acknowledgments}
This work was supported by NNSF of China under Grant No. 10974224
and by the National Basic Research Program of China under Grant No.
2010CB832803.  J.M. would like to thank the Wuhan Institute of
Physics and Mathematics for its hospitality during his visits. The
work of J.M was supported in part by the Australian Research Council
Discovery Project DP-1092620.
\end{acknowledgments}

%%%%%%%%%%%%%%%%%%%%%%%% begin thebibliography  %%%%%%%%%%%%%%%%%%%%%%%%%%
%\bibliography{positron}

\end{document}